\documentclass{emulateapj}
\usepackage{apjfonts}
\usepackage{graphicx}

\usepackage{amsmath}

\shorttitle{Origin of Radially Aligned Magnetic Fields in Young SNRs}
\shortauthors{T. INOUE ET AL.}

\begin{document}

\title{
Origin of Radially Aligned Magnetic Fields in Young Supernova Remnants
}
\author{Tsuyoshi Inoue\altaffilmark{1}, Jiro Shimoda\altaffilmark{1}, Yutaka Ohira\altaffilmark{1}, and Ryo Yamazaki\altaffilmark{1}}
\altaffiltext{1}{Department of Physics and Mathematics, Aoyama-Gakuin University, Sagamihara, Kanagawa 252-5258, Japan; inouety@phys.aoyama.ac.jp}

\begin{abstract}
It has been suggested by radio observations of polarized synchrotron emissions that downstream magnetic fields in some young supernova remnants are oriented radially.
We study magnetic field distribution of turbulent supernova remnant driven by the Richtmyer-Meshkov instability -- in other words, the effect of rippled shock -- by using three-dimensional magnetohydrodynamics simulations.
We find that the induced turbulence has radially biased anisotropic velocity dispersion that leads to a selective amplification of the radial component of the magnetic field.
The Richtmyer-Meshkov instability is induced by the interaction between the shock and upstream density fluctuations.
Future high-resolution polarization observations can distinguish the following candidates responsible for the upstream density fluctuations: (i) inhomogeneity caused by the cascade of large-scale turbulence in the ISM so-called the big-power-law-in-the-sky, (ii) structures generated by the Drury instability in the cosmic-ray modified shock, and (iii) fluctuations induced by the non-linear feedback of the cosmic-ray streaming instability.
\end{abstract}


\section{Introduction}
Supernova remnants (SNRs) are believed to be the sites of galactic cosmic-ray acceleration by the diffusive shock acceleration (DSA; Bell 1978; Blandford \& Ostriker 1978).
However, owing to large ambiguity of magnetic field that plays essential role in the DSA by scattering particles, detailed process of the DSA remains a matter of debate.
Polarization observations of synchrotron emissions can be crucially important, since they have variety of information about the magnetic field in the SNRs.
It has been suggested by radio polarization observations that the magnetic fields in some young SNRs are oriented radially (Dickel et al. 1991 for Tycho's SNR; DeLaney et al. 2002 for Kepler's SNR; Dickel \& Milne 1976, Reynolds \& Gilmore 1993, Reynoso et al. 2013 for SN1006).

Jun \& Norman (1996) studied the Rayleigh-Taylor instability (RTI) at the contact surface in the SNR.
They found that the growth of the RTI causes the radially oriented magnetic field, because the RTI driven turbulence is biased toward the radial component.
Although the RTI provides a promising origin of magnetic field orientation in the SNRs, it can work only in the vicinity of the contact surface (see, however, Schure et al. 2009), while the observations suggest radial orientation even in the region of shocked interstellar medium (ISM) ahead of the contact (e.g., Reynoso et al. 2013).
The Richtmyer-Meshkov instability (RMI) may solve the puzzle of the magnetic field orientation in the SNRs, because the RMI is a Rayleigh-Taylor type instability that is induced by the interaction between the shock and density fluctuations in the ISM, i.e., the RMI can grow even at the downstream of the forward shock.

The role of the RMI in the SNR has recently been studied using magnetohydrodynamics (MHD) simulations.
Giacalone \& Jokipii (2007) and Guo et al. (2012) studied shock propagation of an inhomogeneous ISM with the Kolmogorov-like density power spectrum, and reported that the RMI driven turbulence (in other words, the turbulence driven by the effect of rippled shock) induces small-scale dynamo effect that amplifies the magnetic field.
Inoue et al. (2009; 2010; 2012) and Sano et al. (2012) studied the magnetic field amplification by the RMI driven turbulence due to shock-cloud interaction and discussed its influence on the particle acceleration and high-energy emissions.
However, the influence of the amplified magnetic field by the RMI on the polarization of synchrotron emission has not been studied so far.
In this paper, by using three-dimensional MHD simulations, we study the polarized synchrotron emissions from turbulent SNR driven by the RMI.

\begin{deluxetable}{llllllll} \label{t1}
\tablecaption{Model Parameters}
\tablehead{Model No. & 1 & 2 & 3 & 4 & 5 & 6 & 1f}
\startdata
$\Delta \rho/\langle \rho\rangle_0$ & 0.3 & 0.3 & 0.1 & 0.5 & 0.7 & 0.8 & 0.3 \\
$B_{x,0}$ [$\mu$G] & 0.0 & 3.0 & 0.0 & 0.0 & 0.0 & 0.0 & 0.0 \\
$B_{y,0}$ [$\mu$G] & 3.0 & 0.0 & 3.0 & 3.0 & 3.0 & 3.0 & 3.0 \\
$\langle |\chi_z| \rangle_{P<0.3}$$^{\rm a}$ & 44$^\circ$ & 24$^\circ$ & 72$^\circ$ & 37$^\circ$ & 39$^\circ$ & 38$^\circ$ & 44$^\circ$\\
$\langle |\chi_y| \rangle_{P<0.3}$$^{\rm b}$ & 28$^\circ$ & 23$^\circ$ & 36$^\circ$& 26$^\circ$ & 27$^\circ$ & 27$^\circ$ & 28$^\circ$\\
$\langle P \rangle$$^{\rm c}$ & 0.26 & 0.30 & 0.46 & 0.22 & 0.22 & 0.22 & 0.25\\
$l_{\rm tr}$$^{\rm d}$ [pc] & 0.33 & N/A & N/A & 0.26 & 0.23 & 0.23 & 0.32
\enddata
\tablenotetext{a}{Average polarization angle to the $x$-axis where $P<0.3$ observed along the $z$-axis.}
\tablenotetext{b}{Same as $^{\rm a}$ but observed along the $y$-axis.}
\tablenotetext{c}{Average polarization degree.}
\tablenotetext{d}{Average distance from the shock front at which polarization angle becomes $|\chi|\le 5^{\circ}$.}
\tablenotetext{}{The values of $^{a-d}$ are evaluated at $t=700$ yr, except for Model 1f. Those of Model 1f are evaluated at $t=350$ yr, since the shock velocity is doubled compared to the other models.}
\end{deluxetable}

\section{Setup of Simulations}
We solve the adiabatic ideal MHD equations for the gas with adiabatic index $\gamma=5/3$ and mean molar weight $1.27$, because we are interested in the young SNRs in the late free-expansion phase.
The basic MHD equations are solved using the second-order Godunov-type finite-volume scheme (van Leer 1979) that employs an approximate MHD Riemann solver developed by Sano et al. (1999).
The consistent method of characteristics with the constrained transport technique (Clarke 1996) is used for solving the induction equations that ensures the divergence free constraint.
The integration is done in the conservative fashion, so that we can handle high-Mach-number shock waves precisely.

Since we consider the adiabatic ideal MHD, the results of the simulations are scale-free.
The physical scale is introduced when we fix the scale of density fluctuations given below.
In the following, for intuitive presentation, we express the scale of the system using physical scales.
We prepare a cubic numerical domain of the volume $L_{\rm box}^3=(2\,{\rm pc})^3$ that is composed of 1024$^3$ uniform unit cells.
Because the SNRs that show radially oriented magnetic field seem to be located in the diffuse ISM, we consider so-called the "big-power-law-in-the-sky" (Armstrong et al. 1995) as an origin of the density fluctuations in the preshock ISM.
The fluctuations are given as a superposition of sinusoidal functions with various wave numbers that range $2\pi/L_{\rm box}\le|k|\le256\pi/L_{\rm box}$ and random phases.
The power spectrum of the density fluctuations is given as the isotropic power law: $P_{\rm 1D}(k)\equiv \rho_{k}^2\,k^2\propto k^{-5/3}$ for the above range of $k$, where $\rho_k$ is the Fourier component of the density, which is consistent with the big-power-law-in-the-sky.
Thus, our initial density structure is parameterized by mean density $\langle \rho \rangle_0$ and dispersion $\Delta \rho \equiv (\langle \rho^2 \rangle-\langle \rho \rangle_0^2)^{1/2}$.

We set the mean number density, the initial thermal pressure, and the initial magnetic field strength to be $\langle n \rangle_0=0.5$ cm$^{-3}$, $p/k_{\rm B}=4\times 10^3$ K cm$^{-3}$ and $B_0=3.0\,\mu$G, respectively, which are the typical values in the diffuse ISM (Myers 1978; Beck 2000).
Thus, the initial mean sound speed and Alfv\'en velocity are $\langle c_{\rm s}\rangle_0=9.3$ km s$^{-1}$ and $\langle c_{\rm A}\rangle_0=8.2$ km s$^{-1}$, respectively.
The model parameters (the initial degree of the fluctuation and the initial orientation of magnetic field) are summarized in Table 1.

If we suppose the turbulence in the diffuse ISM is driven by supernovae, the driving-scale of the turbulence and the degree of density fluctuation at the driving-scale would be given as $L_{\rm inj}\sim 100$ pc and $\Delta \rho |_{L_{\rm inj}}/\langle \rho \rangle\sim1$, respectively (e.g., de Avillez \& Breitschwerdt 2007).
In that case, the degree of small-scale density fluctuations due to cascade of the turbulence at the scale $L_{\rm box}=2$ pc is estimated as $\Delta \rho|_{L_{\rm box}}/\langle \rho \rangle\simeq (L_{\rm box}/L_{\rm inj})^{1/3}\simeq 0.27$, where we have used the relation $\Delta \rho^2|_{l}=\int_{1/l}^{\infty} \rho_k^2\,dk^3=l^{2/3}$.
Hence, we regard Models 1 and 2 as fiducial ISM models.

To induce a blast wave that generates a shocked layer, we set a hot plasma of $p_{\rm h}/k_{\rm B}=2\times 10^8$ K cm$^{-3}$, $n_{\rm h}=0.05$ cm$^{-3}$ and $B_{\rm h}=3.0$ $\mu$G at the $x=0$ boundary plane.
According to the solution of the Riemann problem, when the preshock density takes uniform value of $n=0.5$ cm$^{-3}$, such a hot gas induces a shock wave of $v_{\rm sh}=1795$ km s$^{-1}$, indicating that mean shock velocity induced by the hot gas is $\langle v_{\rm sh}\rangle\simeq 1800$ km s$^{-1}$.
We use the periodic boundary conditions for the $x$-$y$ and the $x$-$z$ boundary planes, and we assume free boundary condition at the $x=L_{\rm box}$ boundary plane.
In the young SNRs such as Tycho, Kepler, and SN1006, the shock speed does not vary substantially, since the SNRs would be in the late free-expansion phase.
In addition, our numerical domain ($L_{\rm box}= 2$ pc) is smaller than the curvature of the young SNRs ($\sim10$ pc).
Thus, the dynamical effect of the curvature can be omitted.
To study the effect of the shock strength, we also perform the run named Model 1f that is similar to Model 1 with twice the shock speed ($\langle v_{\rm sh}\rangle\simeq3600$ km s$^{-1}$) by enhancing $p_{\rm h}$ by a factor of four.

\begin{figure}[t]
\epsscale{1.1}
\plotone{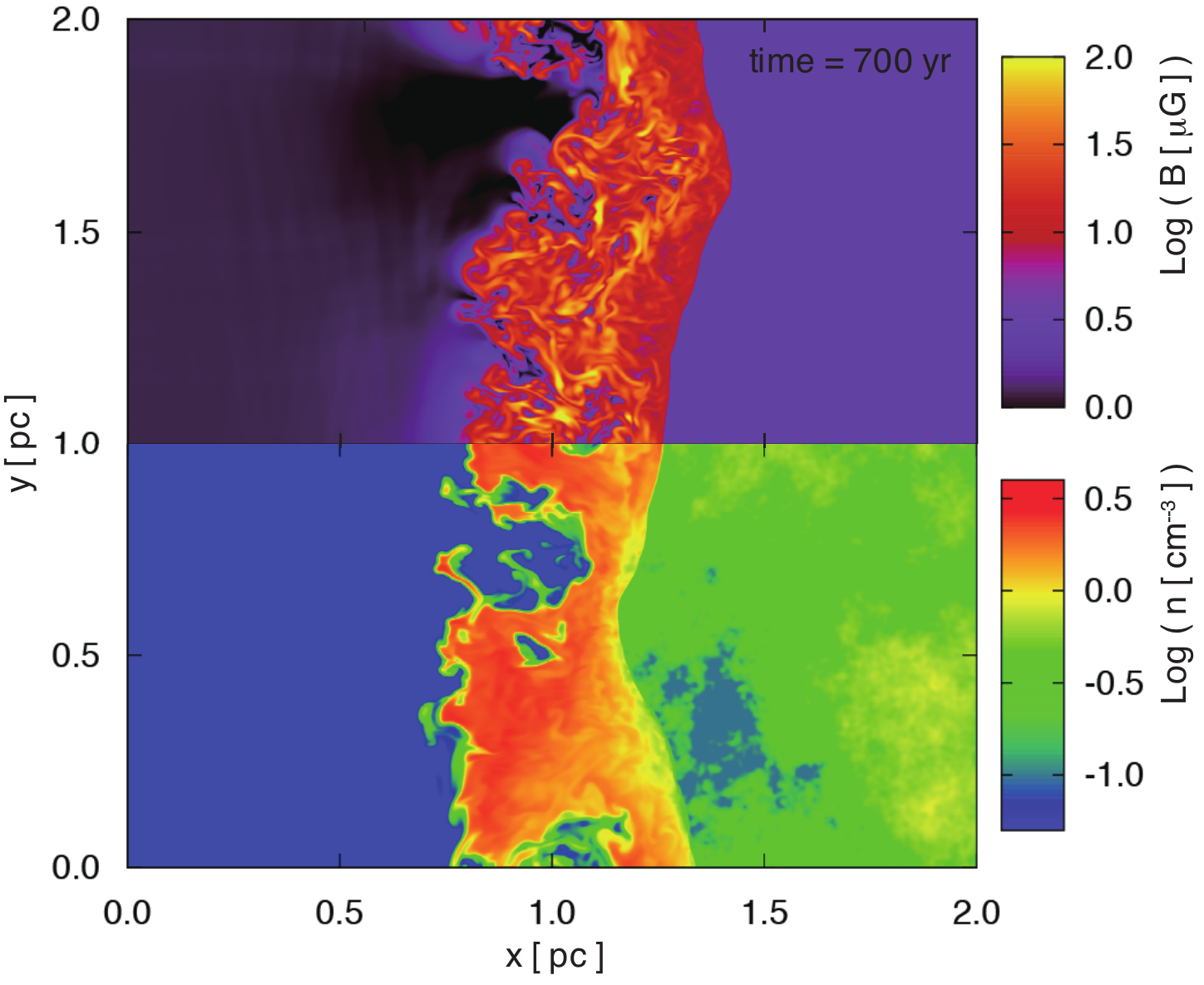}
\caption{\label{f1}
Two-dimensional slice of the magnetic field strength ({\it upper half}) and the number density ({\it lower half}) of the result of Model 1 at $t=700$ yr and $z=0$ pc plane.
}\end{figure}

\begin{figure}[t]
\epsscale{1.7}
\plotone{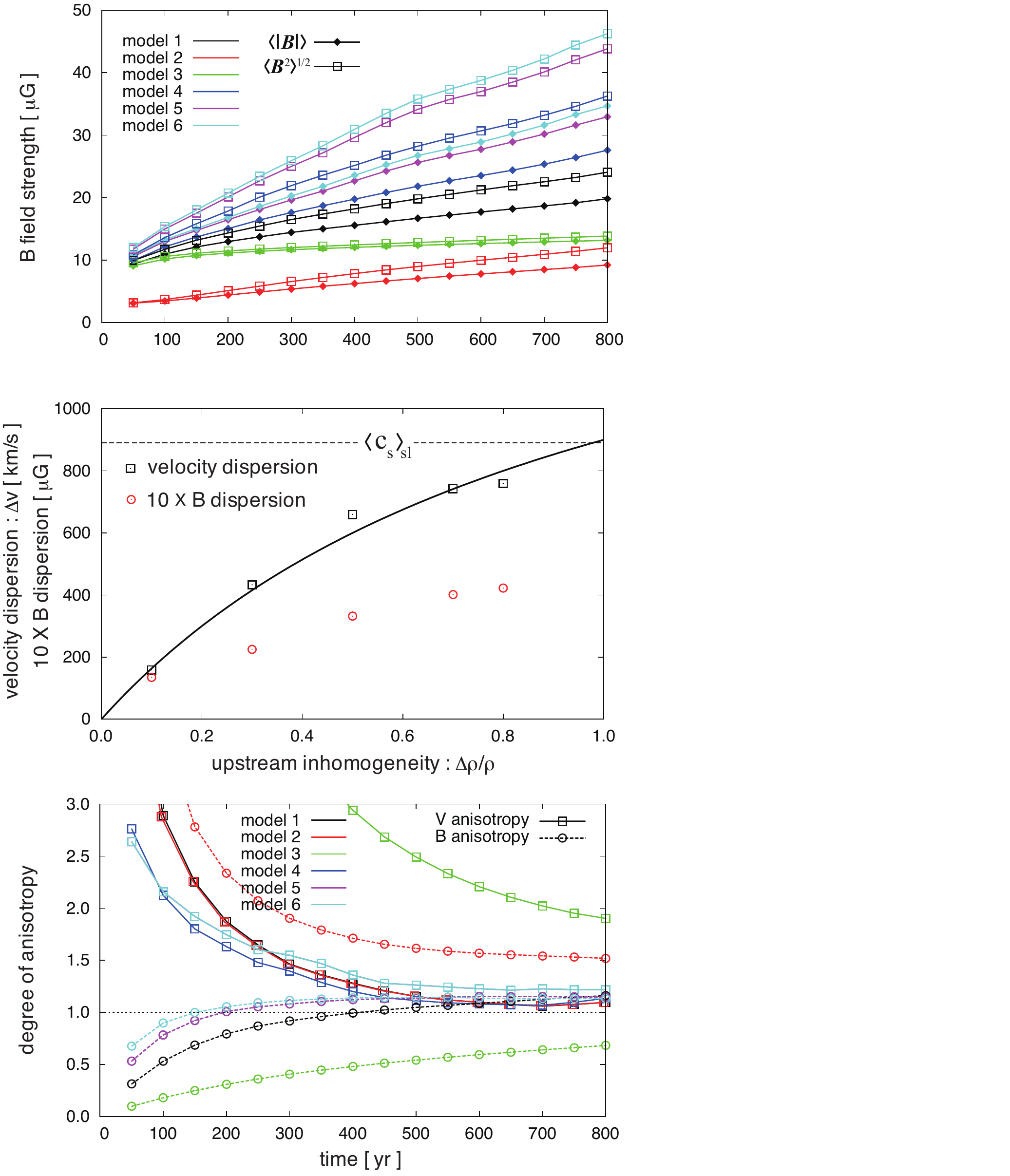}
\caption{\label{f2}
Top: Temporal evolution of the average magnetic field strength $\langle |{\bf B}| \rangle_{\rm sl}$ and the dispersion $\langle {\bf B}^2 \rangle_{\rm sl}^{1/2}$ in the shocked layer.
Mid: Velocity dispersion $\Delta v$ and the dispersion of magnetic field strength $\Delta B$ at $t=700$ yr for the perpendicular shock cases as functions of the upstream density inhomogeneity $\Delta \rho/\langle \rho \rangle_0$.
Solid line shows a theoretical estimation of the velocity dispersion given by eq.~(\ref{vRMI}), and dashed line indicates mean postshock sound speed ($\langle c_{\rm s}\rangle_{\rm sl}$).
Bottom: Ratio of $v_x$ dispersion to the average value of $v_y$ and $v_z$ dispersions [$r_{v}\equiv 2\,\Delta v_x/(\Delta v_y+\Delta v_z)$] ({\it solid}) and ratio of the average strength of $B_x$ to the average value of $B_y$ and $B_z$ strengths [$r_B\equiv 2\langle |B_x|\rangle_{\rm sl}/(\langle |B_y|\rangle_{\rm sl}+\langle |B_z|\rangle_{\rm sl})$] ({\it dotted}).
}\end{figure}

\begin{figure*}[t]
\epsscale{1.1}
\plotone{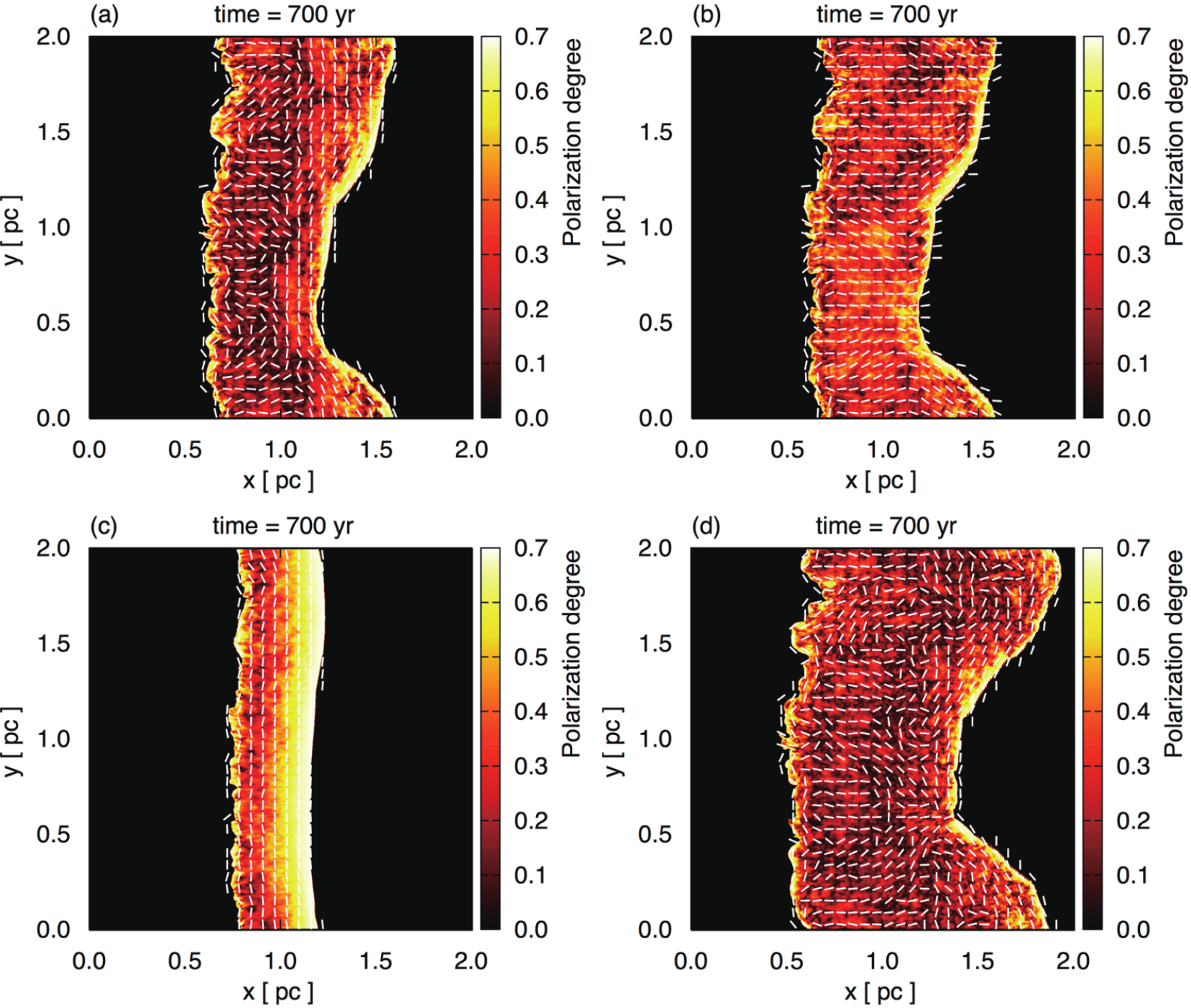}
\caption{\label{f3}
Structures of polarization degree $\sqrt{Q^2+U^2}/I$.
White bars indicate the magnetic field orientation vectors derived from the polarization angle $\chi=\tan^{-1}(U/Q)/2$.
Panel (a)-(d) correspond to the results of Models 1-4, respectively.
The choice of the l.o.s. along $z$-axis corresponds to the case in which expectation value of the polarization angle $\langle|\chi|\rangle$ takes maximum.
}\end{figure*}

\section{Results of Simulations and Synthetic Observations}
In Figure~\ref{f1}, we show a two-dimensional slice of the magnetic field strength ({\it upper half}) and the number density ({\it lower half}) of the result of Model 1 at $t=700$ yr.
Interaction between the blast wave shock and the density fluctuations results in an induction of turbulence, where the magnetic field is amplified beyond the shock compression value of $|B|=12$ $\mu$G for the perpendicular shocks propagating into uniform medium.
Top panel of Figure~\ref{f2} shows evolution of the average magnetic field strength $\langle |{\bf B}| \rangle_{\rm sl}$ and the dispersion $\langle {\bf B}^2 \rangle_{\rm sl}^{1/2}$ in the shocked layer.
Here $\langle \cdots \rangle_{\rm sl}$ represents the mean value in the shocked region where $p/k_{\rm B}>10^5$ K cm$^{-3}$ and $n>0.3$ cm$^{-3}$ (the former and latter conditions exclude the preshock ISM and the hot ejecta component, respectively).
One can find more detailed descriptions about the RMI driven turbulence (or the effect of rippled shock in another expression) and resulting dynamo effect in many literatures (e.g., Giacalone \& Jokipii 2007; Inoue et al. 2012; Fraschetti 2013).
The middle panel of Figure~\ref{f2} shows, for the perpendicular shock models, the velocity dispersion in the shocked layer $\Delta v\equiv (\sum_i \langle v_i^2 \rangle_{\rm sl}-\langle v_i \rangle_{\rm sl}^2)^{1/2}$ and the dispersion of magnetic field strength $\Delta B$ at
$t=700$ yr as functions of the preshock density dispersion.
If we apply the result of a simple linear analysis by Richtmyer (1960), the velocity dispersion of shocked density fluctuations that is essentially given by the growth velocity of the RMI can be written as
\begin{equation}\label{vRMI}
\Delta v\simeq v_{\rm RMI}\simeq A\,\langle v_{\rm sh}\rangle \,(1-\eta),
\end{equation}
where $A\simeq (\Delta \rho/\rho)/(1+\Delta \rho/\rho)$ corresponds to the Atwood number and $\eta$ is the ellipticity of the fluctuations that is zero for the isotropic inhomogeneity (see Nishihara et al. 2010; Mikaelian 1996; Inoue 2012 for more sophisticated analyses).
Eq.~(\ref{vRMI}) with $\langle v_{\rm sh}\rangle=1800$ km s$^{-1}$ and $\eta=0$ is plotted in the middle panel of Figure~\ref{f2} as a solid line.
We can see the good agreement between the theoretical estimation and simulations, and it is clear that $\Delta v$ saturates for large $\Delta \rho/\rho$.
 
To calculate the polarization degree and angle of synchrotron emission from the shocked layer, we use the formulae following Clarke et al. (1989):
\begin{eqnarray}
i(s) &=& K(\nu)\,\nu^{-\alpha}\,\{|{\bf B}(s)|\,\sin\psi(s)\}^{1+\alpha},\\ 
I &=& \int_{\rm l.o.s.}i(s)\,ds,\\
Q &=&  \int_{\rm l.o.s.}f_0\,i(s)\,\cos[2\phi(s)]\,ds,\\
U &=&  \int_{\rm l.o.s.}f_0\,i(s)\,\sin[2\phi(s)]\,ds,
\end{eqnarray}
where $K(\nu)$ is a function depending on the density of relativistic electrons, $\alpha=(p-1)/2$ ($p=2$ is the spectral index of the relativistic electrons by the DSA), $f_0=(\alpha+1)/(\alpha+5/3)$ is a local linear polarization degree for isotropic electrons, $\psi$ is an angle between local magnetic field and the line of sight, and $\phi$ is a position angle of the local magnetic field projected onto the plane of the sky and the $x$-axis.
In the following, we assume $\alpha=0.5$ and the relativistic electrons are uniformly distributed in the shocked region ($K=$const.).

Figure~\ref{f3} shows the structure of polarization degree $P\equiv \sqrt{Q^2+U^2}/I$ (colors) and the magnetic field orientation vectors derived from the polarization angle $\chi=\tan^{-1}(U/Q)/2$ (bars) observed along the $z$-axis.
Panels (a)-(d) correspond to the results of Model 1-4, respectively.
We see that the orientation angle to the $x$-axis $\chi$ decrease with distance from the shock, even in the cases of the perpendicular shock.
In particular, substantial fraction of the orientation vectors are aligned almost radially ($x$-direction) in the results of $\Delta \rho/\rho\ge0.3$ (panels [a] and [d]).
Note that the l.o.s. direction along $z$-axis corresponds to the situation in which the expectation value of the polarization angle to the $x$-axis $\langle|\chi|\rangle$ is the largest in any other l.o.s. direction.
In Table~1, we list the average polarization angles for the cases of the l.o.s. along $z$-axis ($\langle |\chi_z| \rangle_{P<0.3}$) and the l.o.s. along $y$-axis ($\langle |\chi_z| \rangle_{P<0.3}$), in which the average is taken in the region of $P<0.3$ where the effect of turbulence is active.
In the models of $\Delta\rho/\rho\ge0.3$, both angles are smaller than that of the isotropic case: $45^{\circ}$.

The mechanism for this transition of the orientation vector can be explained as follows:
In the bottom panel of Figure~\ref{f2}, we plot the ratio of $v_x$ dispersion to the average value of $v_y$ and $v_z$ dispersions [$r_{v}\equiv 2\,\Delta v_x/(\Delta v_y+\Delta v_z)$] ({\it solid}) and the ratio of the average strength of $B_x$ to the average value of $B_y$ and $B_z$ strengths [$r_B\equiv 2\langle |B_x|\rangle_{\rm sl}/(\langle |B_y|\rangle_{\rm sl}+\langle |B_z|\rangle_{\rm sl})$] ({\it dotted}).
We see that the velocity dispersion is anisotropic that is biased toward $x$-component.
This indicates that the turbulence, which stretches and amplifies magnetic field lines, works selectively for the $x$-component magnetic field.
As a result, the magnetic field orientation shifts to the $x$-direction.
Note that, since the isotropic component of the magnetic field is averaged out by the l.o.s. integration, the radial orientation of the magnetic field appears even if the anisotropy of the magnetic field strength is weak.
This mechanism is essentially the same as the RTI case at the contact surface studied by Jun \& Norman (1996), and possibly explains small dispersion of dust emission/absorption polarization angles in molecular clouds as discussed in Inoue \& Inutsuka (2012).
As for Model 3 ($\Delta \rho/\rho=0.1$), the anisotropy of magnetic field strength $r_B$ do not exceed unity in 800 yr evolution due to the smaller velocity dispersion.
This results in the insufficient polarization transition in the panel (c).

The mean polarization degrees $\langle P \rangle$ at $t=700$ yr are exhibited in Table 1.
The polarization degree of $\langle P \rangle\simeq 20$-$30$\% in the shocked layer for the cases of $\Delta \rho/\rho\ge 0.3$ agrees with the radio polarization observations of young SNRs (Dickel et al. 1991; DeLaney et al. 2002; Reynolds \& Gilmore 1993; Reynoso et al. 2013).
These lower polarization degrees $\langle P \rangle$ compared to the homogeneous magnetic field case of $P\simeq 70$\% for the standard DSA electron spectrum is clearly attributed to the turbulent structure of the magnetic field.

Because the transition of polarization angle is caused by the anisotropic turbulent dynamo effect, larger velocity dispersion of turbulence results in a smaller transition length of the polarization angle.
The transition lengths of the polarization angle for the perpendicular shock cases are $l_{\rm tr}=0.33$ pc for Model 1 ($\Delta\rho/\rho=0.3$), 0.25 pc for Model 4 ($\Delta\rho/\rho=0.5$), 0.23 pc for Model 5 ($\Delta\rho/\rho=0.7$), and 0.23 pc for Model 6 ($\Delta\rho/\rho=0.8$).
Here the transition length is defined as the average distance from the shock front at which polarization angle becomes $|\chi|\le 5^{\circ}$.
Note that the transition lengths are converged at $t\simeq 500$ yr within $\simeq20$\% deviations from the above values.
We also confirmed that the transition length are converged within $\simeq20$\% deviations, even if we employed $|\chi|\le 10^{\circ}$.

In the present set of problem, the distribution of magnetic field that determines $l_{\rm tr},\,\langle P\rangle$, and $\langle \chi\rangle$ is passive to the turbulence.
The velocity dispersion of the turbulence is described by eq.~(\ref{vRMI}), and the characteristic scale of the turbulent modification of the magnetic field can be written as (Richtmyer 1960; Sano et al. 2012)
\begin{equation}\label{lRMI}
 l_{\rm RMI}\sim v_{\rm d}\,t_{\rm RMI}\simeq v_{\rm d}\,l_{\Delta \rho}/\{A\,v_{\rm sh}\,(1-\eta)\}\simeq l_{\Delta \rho}/(r_{\rm c}\,A),
 \end{equation}
where $v_{\rm d}$ is the downstream velocity and $r_{\rm c}$ is the shock compression ratio.
It is widely known that $r_{\rm c}$ converges quickly to the value of four for the high Mach number shock in the monoatomic gas.
Thus, eq.~(\ref{lRMI}) indicates that the modification scale of the magnetic field is insensitive to the Mach number or the shock velocity.
The result of Model 1f, which is similar to Model 1 except the shock velocity is doubled ($\langle v_{\rm sh}\rangle\simeq 3600$ km s$^{-1}$), confirms the insensitivity of $l_{\rm tr},\,\langle P\rangle$, and $\langle \chi\rangle$ to the shock velocity (see, Table 1).

\section{Discussion}
Can the RMI driven turbulence account for the observed radially oriented magnetic fields in the young SNRs?
Recent radio polarization observation of SN1006 has shown that the magnetic field orientations in southwest (SW) and southeast (SE) regions seem to be shifted from perpendicular (to the shock normal) to parallel (Reynoso et al. 2013).
Interestingly enough, these two regions show different features:
In the SW region, where $P\sim 20\%$ and non-thermal x-ray emission is bright, the transition length of the polarization angle is apparently $\lesssim 10''$ that corresponds to $l_{\rm tr}\lesssim 0.1$ pc for the distance $d=2.18$ kpc (see, Figure~3 of Reynoso et al. 2013).
On the other hand, in the SE, the transition length is apparently $\gtrsim 30''$ ($l_{\rm tr}\gtrsim 0.3$) pc, where the polarization degree is high $P\sim 70\%$ and the non-thermal x-ray emission is faint.

As we have discussed in \S 2 that the upstream density dispersion of $\Delta \rho/\rho \simeq 0.3$ (for $L_{\rm box}=2$ pc) may be the standard fluctuation amplitude in the diffuse ISM.
If so, the transition length of the polarization angle is $l_{\rm tr}\simeq 0.3$ pc (Table 1) that is comparable to that of the SE region of SN1006.
The averaged polarization degree of our fiducial model (Model 1) $\langle P \rangle \sim 30\%$ is much smaller than the SE value of $P\sim 70\%$.
However, the polarization degree of the SE region is measured only near the rim due to the faint emission.
Figure~\ref{f3} indicates that the polarization degree at the rim of the shock is also large ($P\gtrsim 0.5$) in our model.
Therefore, Model 1 is consistent with the SE region of SN1006.

As for the SW region, the transition length of the fiducial model ($l_{\rm tr}\simeq 0.3$ pc) is larger than that measured by the observation($\lesssim 0.1$ pc).
In addition, even at the rim of the shell, the polarization degree of the SW is low ($P\sim 20\%$).
Even if the amplitude of the fluctuations is much larger than the fiducial model ($\Delta \rho/\rho>0.3$), the velocity dispersion of induced turbulence can be at most as large as the cases of Models 4-6, because the induced velocity dispersion saturates at $\Delta \rho/\rho \simeq 0.6$ (see, mid panel of Figure~\ref{f2}).
This indicates that the larger density inhomogeneity does not substantially shorten the transition length (see, Table 1) and the thickness of the rim with $P>0.5$ (see, Figure~\ref{f3}).

The observed transition length of the southwest region of SN1006 can be much smaller than 0.1 pc ($10''$), since it is comparable to the scale of beam spread.
In that case, we have interesting options to address the scale discrepancy between the theory and the observation of the SW region as follows:
As we have mentioned in \S 2, since our mechanism is based on the scale-free ideal MHD phenomenon, we can expect smaller scale polarization shift, if there are small-scale fluctuations with sufficient amplitude.
In other words, if we re-normalize the scale of our numerical domain to be 1/10th of our choice, the same result with the 1/10th of spatial and temporal scales without changing other variables is obtained.
We can reasonably expect large-amplitude small-scale density fluctuations around the young SNR shock, if the cosmic-ray acceleration is efficient.
For instance, the Drury instability can be an origin of the large-amplitude small-scale density fluctuations, which amplifies preexistent sound waves in the precursor of the cosmic-ray modified shock (Drury \& Falle 1986).
Also, recent particle-in-cell simulations have shown that the cosmic-ray streaming instability (e.g, Bell 2004) causes density fluctuations in the upstream region in its non-linear stage (Riquelme \& Spitkovsky 2009; Niemiec et al. 2008; Ohira et al. 2009).
In addition, the density fluctuations could be produced by pickup ions originated in hydrogen atoms (Ohira \& Takahara 2010; Ohira 2012).

In the case of the Drury instability, the most unstable scales of the acoustic waves are (Malkov et al. 2010)
\begin{equation}
l_{\rm D}\lesssim 4\pi\rho_{\rm up}\,c_{\rm s,up}^2/(\partial_x p_{\rm CR}),
\end{equation}
where $\partial_x\,p_{\rm CR}$ is the cosmic-rays pressure gradient in the shock precursor.
As for the non-linear stage of the cosmic-ray streaming instability, the typical scale is given by (Bell 2004)
\begin{equation}
l_{\rm cs}\simeq B_{\rm up}\,c/(2\,j_{\rm CR}),
\end{equation}
where $j_{\rm CR}= e\,n_{\rm CR}\,v_{\rm sh}$ is the cosmic-ray current density and $B_{\rm up}$ is the upstream magnetic field strength amplified by the instability that can be measured via the thickness of thin synchrotron shell.
Since the typical scale of the RMI given in eq.~(\ref{lRMI}) roughly gives the transition length ($l_{\rm tr}\sim l_{\rm RMI}$), once the scale of the transition length is measured with sufficient resolution, we can recover the scale of the Drury instability or the cosmic-ray streaming instability by substituting $l_{\rm D}$ or $l_{\rm cs}$ for $l_{\Delta \rho}$.
This method potentially enable us to measure the cosmic-ray pressure gradient: $\partial_x p_{\rm CR}\sim 4\pi\,v_{\rm d}\,\rho_{\rm up}\,c_{\rm s,up}^2/(A\,v_{\rm sh}\,l_{\rm tr})$ or the cosmic-ray number density: $n_{\rm CR}\sim v_{\rm d}\,B_{\rm up}\,c/(2\,A\,e\,v_{\rm sh}^2\,l_{\rm tr})$ by observations.

\acknowledgments
Numerical computations were carried out on XC30 system at the Center for Computational Astrophysics (CfCA) of National Astronomical Observatory of Japan and K computer at the RIKEN Advanced Institute for Computational Science (No. hp120087).
This work is supported by Grant-in-aids from the Ministry of Education, Culture, Sports, Science, and Technology (MEXT) of Japan, No. 23740154 (T. I.) and No. 248344 (Y. O.).
T. I. and R. Y. deeply appreciate Research Institute, Aoyama-Gakuin University for helping our research by the fund.

\end{document}